\documentclass[aps, prx, superscriptaddress, twocolumn, amsfonts, amsmath, amssymb]{revtex4-2}
\usepackage{graphicx}
\usepackage{subfigure}
\usepackage{xcolor}
\usepackage{physics}
\usepackage{enumerate}
\usepackage{siunitx}
\newcommand{\phiext}{\phi_{\mathrm{ext}}}
\newcommand{\phiexta}{\phi_{\mathrm{ext,1}}}
\newcommand{\phiextb}{\phi_{\mathrm{ext,2}}}
\newcommand{\rettelse}[1]{#1}

\begin{document}
\title{Fast universal control of a flux qubit via exponentially tunable wave-function overlap}

\author{Svend Kr{\o}jer}
\affiliation{Center for Quantum Devices, Niels Bohr Institute, University of Copenhagen, DK-2100 Copenhagen, Denmark}
\author{Anders Enevold Dahl}
\affiliation{Center for Quantum Devices, Niels Bohr Institute, University of Copenhagen, DK-2100 Copenhagen, Denmark}
\author{Kasper Sangild Christensen}
\affiliation{Center for Quantum Devices, Niels Bohr Institute, University of Copenhagen, DK-2100 Copenhagen, Denmark}
\affiliation{Department of Physics and Astronomy, Aarhus University, Ny Munkegade 120, 8000 Aarhus C, Denmark}
\author{Morten Kjaergaard}
\affiliation{Center for Quantum Devices, Niels Bohr Institute, University of Copenhagen, DK-2100 Copenhagen, Denmark}
\author{Karsten Flensberg}
\affiliation{Center for Quantum Devices, Niels Bohr Institute, University of Copenhagen, DK-2100 Copenhagen, Denmark}

\date{\today}

\begin{abstract}
Fast, high fidelity control and readout of protected superconducting qubits are fundamentally challenging due to their inherent insensitivity. We propose a flux qubit variation which enjoys a tunable level of protection against relaxation to resolve this outstanding issue. Our qubit design, the double-shunted flux qubit (DSFQ), realizes a generic double-well potential through its three junction ring geometry. One of the junctions is tunable, making it possible to control the barrier height and thus the level of protection. We analyze single- and two-qubit gate operations that rely on lowering the barrier. We show that this is a viable method that results in high fidelity gates as the non-computational states are not occupied during operations. Further, we show how the effective coupling to a readout resonator can be controlled by adjusting the externally applied flux while the DSFQ is protected from decaying into the readout resonator. Finally, we also study a double-loop gradiometric version of the DSFQ which is exponentially insensitive to variations in the global magnetic field, even when the loop areas are non-identical.
\end{abstract}

\maketitle

\section{Introduction}

Qubits based on superconducting junctions form a promising platform for quantum computation (QC) architectures \cite{kjaergaard_superconducting_2020, gyenis_moving_2021, siddiqi_engineering_2021}. In order to scale up fault-tolerant QC, it is crucial that
gate and readout \rettelse{in}fidelities \rettelse{must be lower than the} threshold for quantum error correction (QEC), which for the surface code is about 1\% \cite{Dennis_quantum_memory_2002, Fowler_Surface_Codes_2012}. A number of experiments using transmon-based multi-qubit chips have demonstrated surface code QEC close to the threshold \cite{zhao_realization_2022, krinner_realizing_2022, Acharya2023}. 

To go beyond the capabilities of contemporary transmon-based architectures, a number of $T_1$-protected qubit designs have appeared \cite{brooks_protected_2013, Earnest_heavy_fluxonium_2018,larsen_parity-protected_2020,kalashnikov_bifluxon_2020, gyenis_moving_2021}. The general idea of a $T_1$-protected superconducting qubit is that the computational states are localized in different quantum wells, leading to exponentially suppressed noise-induced transitions, enhancing the relaxation time significantly \cite{gyenis_moving_2021}. Additionally, the double-well potential realizes low-frequency qubits resulting in less sensitivity to dielectric loss and Ohmic noise channels \cite{groszkowski_coherence_2018, Nguyen2022blueprint}. 

In the flux qubit modality, this kind of double-well protection can be reached by biasing the superconducting loop with an external flux close to half a flux quantum \cite{orlando_superconducting_1999, manucharyan_fluxonium_2009}. Here, the low-energy computational states corresponds to supercurrent flowing in opposite directions in the loop. At a bias of half a flux quantum, the fluxon states are degenerate up to the exponentially small \rettelse{splitting due to overlap of the evanescent part of the wave functions} across the barrier separating the two wells. \rettelse{Below we refer to this small splitting as the wave-function overlap.} The fluxon states are sensitive to the external magnetic flux as it picks out a preferred current direction and determines the energy splitting. The strong flux dependence leads to a linear sensitivity of the qubit frequency to flux noise, causing dephasing of the qubit and limiting coherence \cite{mooij_josephson_1999, Earnest_heavy_fluxonium_2018}. 

Despite the enhanced relaxation time of low-frequency qubits (e.g.\ heavy fluxonium \cite{Earnest_heavy_fluxonium_2018, zhang_universal_2021}, $0-\pi$ qubit \cite{brooks_protected_2013, gyenis_experimental_2021}, etc.), a general disadvantage is that gate times typically also increase due to the vanishing wave-function overlap of the computational states. One way of circumventing this limitation is to use higher lying non-computational states \cite{Earnest_heavy_fluxonium_2018, gyenis_moving_2021, gyenis_experimental_2021}. In this manner, single and two qubit gates can be activated through multi-tone driving \cite{abdelhafez_universal_2020}. The downside of \rettelse{such an} approach, however, is that the momentary occupancy of the non-computational states leads to increased decoherence, limiting gate fidelities \cite{ficheux_fast_2021}. Another possibility is to rely on diabatic single qubit control \cite{zhang_universal_2021}.

In this paper, we explore an alternative approach to perform gates on $T_1$-protected qubits that rely on adiabatically adjusting the level of protection by lowering the barrier between the two wells. We propose a qubit design, the double-shunted flux qubit (DSFQ), which aims to be a relatively simple modification of a flux qubit with exponentially tunable wave-function overlap. The DSFQ is related to the persistent current flux qubit (PCFQ) \cite{orlando_superconducting_1999, mooij_josephson_1999} and the capacitively shunted flux qubit (CSFQ) \cite{yan_flux_2016} as they all share the same circuit layout of three Josephson junctions (JJs) connected in a loop, see Fig.\ \ref{fig:fig1}. While the PCFQ realizes a large $E_J/E_C$ via three large junctions, the CSFQ uses smaller junctions with one large capacitive shunt such that one mode is heavy (large $E_J/E_C$) and one mode is light (smaller $E_J/E_C$). The DSFQ finds the middle ground between these designs by using small junctions and two large capacitive shunts such that both modes are heavy, similar to the PCFQ. Since both modes are heavy, the lowest energy wave functions are localized in separate wells, protecting the qubit from relaxation. Other designs, namely the super-semi $\cos(2\phi)$ qubit and the bifluxon, have successfully shown an order-of-magnitude improvement of the relaxation time in the protected regime \cite{larsen_parity-protected_2020, kalashnikov_bifluxon_2020}. However, both qubits are challenging to fabricate and tune to the ideal regime and two-qubit gates have not yet been realized \cite{gyenis_moving_2021}. The DSFQ offers a comparatively simple platform for studying universal gate sets for qubits with variable wave-function overlap. In addition to the universal gate scheme, we also propose a noise-insensitive readout method for the DSFQ.

We imagine tuning the barrier height by a tunable junction, implemented either in a SQUID-loop as in previous PCFQ experiments \cite{paauw_tuning_2009, Zhu_tunable_gap_flux_qubit_2010, Gustavsson_tunable_tunnel_2011, schwarz_gradiometric_2013} or in a hybrid version where the tunable junction is a superconductor-semiconductor-superconductor junction. This type of junction has been demonstrated earlier to be stable and having coherence times longer than the anticipated gate times \cite{larsen_semiconductor-nanowire-based_2015, Casparis_Gatemon_2016, casparis_superconducting_2018, larsen_parity-protected_2020, Hertel_Gate_transmon_2022}. \rettelse{However, we note that the coherence times for the  semiconductor-based junctions are still shorter than the more standard insulator-barrier junctions. The physics of this is still not understood and the coherence times could improve with future devices \cite{aguado_perspective_2020}.}

We calculate the coherence properties of the DSFQ and discuss the flux-noise sensitivity. In order to reduce the flux dephasing, we propose a double-loop gradiometric version of the DSFQ which gives exponential protection against global flux noise. Gradiometric qubit designs have been proposed previously but rely on identical areas in the two loops \cite{paauw_tuning_2009, schwarz_gradiometric_2013, gusenkova_operating_2022}. We show that small area variations can be compensated for by adjusting the tunable junction without introducing sensitivity to the junction control line. The main focus of our study is a set of one- and two-qubit gates where the idea is to tune the qubit out of the protected regime by adiabatically lowering the barrier between the two wells and thereby hybridize the computational states. Two-qubit gates can be performed by simultaneously lowering the barriers for two capacitively coupled DSFQ's while single qubit gates require a fast single-tone microwave pulse in the an intermediate regime. Advantages of \rettelse{variable-protection gates} are that fast-decaying non-computational states do not participate in gate operations and that two-qubit interactions can be turned off with exponential on/off ratio while maintaining the ability to perform one-qubit gates. Finally, we show how the effective coupling to a readout resonator can be adjusted with a simple flux control of the qubit, leading to an order-of-magnitude on/off ratio while decay to the readout resonator is suppressed.

\section{The double-shunted flux qubit}
\label{sec:theory}

\begin{figure}[t]
    \centering
    \includegraphics{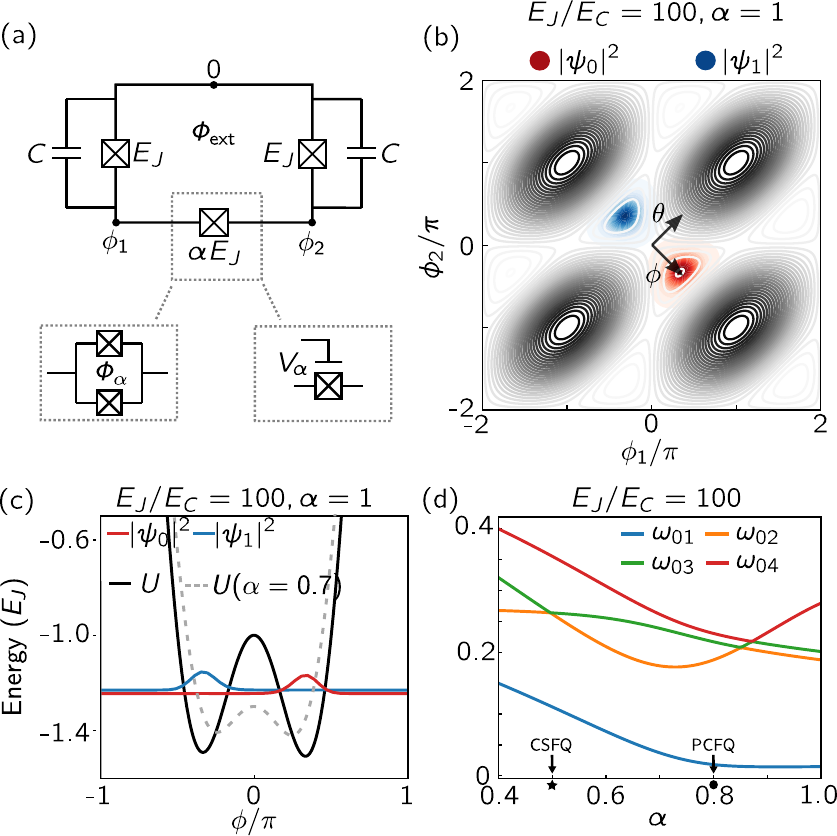}
    \caption{(a) Circuit layout for the DSFQ with a variable junction by either a SQUID or gate voltage tunable nanowire junction. (b) Potential landscape of the DSFQ with the two lowest energy eigenstates shown in red and blue with $E_J/E_C = 100,\alpha = 1$ and $\phiext = 0.997\pi$. (c) One dimensional cut of (b) along $\phi = (\phi_1 - \phi_2)/2$ with wave functions showing their exponential separation \rettelse{at $\alpha = 1$. The potential at $\alpha = 0.7$ is shown in gray dashed}. (d) Energy splitting of the qubit as a function of the barrier height controlled by $\alpha$. The value of $\alpha$ corresponding to the CSFQ/PCFQ is indicated with a star/bullet ($\alpha = 0.5/0.8$). \rettelse{Energies are in units of the Josephson energy, $E_J$.}}
    \label{fig:fig1}
\end{figure}

We consider a system of three Josephson junctions connected in a ring. The circuit is illustrated in Fig.~\ref{fig:fig1}(a) where the Josephson energy of the tunable junction is denoted by $\alpha E_J$. The two other junctions have Josephson energy $E_J$, but they do not have to be identical for our proposal to work. In the phase variables $\phi=(\phi_1-\phi_2)/2$ and $\theta = (\phi_1+\phi_2)/2$, the potential energy of the qubit is thus given by
\begin{equation}\label{Hpot}
  H_J=
  -2 E_J \cos(\phi)\cos(\theta)
  -\alpha E_J\cos(2\phi+\phiext),
\end{equation}
where $\phiext=2\pi\Phi/\Phi_0$ and $\Phi$ is the flux through the loop, controlled by an external magnetic field whose value is typically set \rettelse{to $\phiext = 0.997 \pi$ unless other stated}. At $\alpha=0.5$, the barrier is completely lowered, making the potential along the $\phi$-direction approximately quartic as for the CSFQ \cite{yan_flux_2016}. At a value of $\alpha=0.8$, the barrier is significant and the potential of the PCFQ \cite{orlando_superconducting_1999,mooij_josephson_1999} is recovered. Controlling the barrier height of the DSFQ through $\alpha$ thus interpolates between the PCFQ and the CSFQ. Note that \rettelse{in the flux-tunable PCFQ, the barrier height can be controlled via an external flux in a slightly different geometry} \cite{paauw_tuning_2009, Gustavsson_tunable_tunnel_2011, schwarz_gradiometric_2013}.

The charging energy is determined by the capacitances $C$ shown in Fig.~\ref{fig:fig1}(a) and gives rise to the kinetic energy \cite{rasmussen_superconducting_2021}
\begin{equation}\label{Hkin}
  H_C=2E_C\left(-i\partial_{\phi}-n_{g\phi}\right)^2 +2E_C\left(-i\partial_{\theta}-n_{g\theta}\right)^2.
\end{equation}
Here we have included offset charges $n_{g\phi}$ and $n_{g\theta}$ (the $4E_C$ typically found as the prefactor is reduced due to the change of variables $n_{\theta/\phi}=n_1\pm n_2$). The qubit will be operated in the regime of small $E_C = e^2/2C$ (i.e., both $\phi$ and $\theta$ being heavy modes). \rettelse{Realistically, the }Josephson capacitances are \rettelse{about two orders of magnitude smaller than the large shunting capacitances and thus merely renormalizes $E_C$ without affecting the results presented in this work.}

The potential landscape and the ground-state wave functions are shown in Fig.~\ref{fig:fig1}(b) in the heavy-modes regime ($E_J/E_C=100$). The external flux is tuned to a value close to half a flux quantum. The two wave function shown in red and blue ($\psi_0$ and $\psi_1$) are clearly well separated and localized in the two wells. They represent the qubit states $\ket{0}$ and $\ket{1}$. \rettelse{The state separation} is most easily seen in  Fig.~\ref{fig:fig1}(c) which is a cut along the $\phi$-direction. Due to their separation, the tunneling between the two wells is suppressed. It results in a small qubit splitting near $\alpha=1$ determined by the external flux and also a large anharmonicity, see Fig.~\ref{fig:fig1}(d). Lowering the barrier by reducing $\alpha$, increases the qubit frequency and decreases the anharmonicity  $\alpha_{\text{an}}=(\omega_{02} - \omega_{01}) - \omega_{01}$, as the states hybridize \rettelse{and change significantly}. This fact is used below to perform fast gates by lowering the value of $\alpha$ to $\alpha\approx 0.7$ where the logical states partially overlap.

\begin{figure*}[ht]
    \centering
    \includegraphics{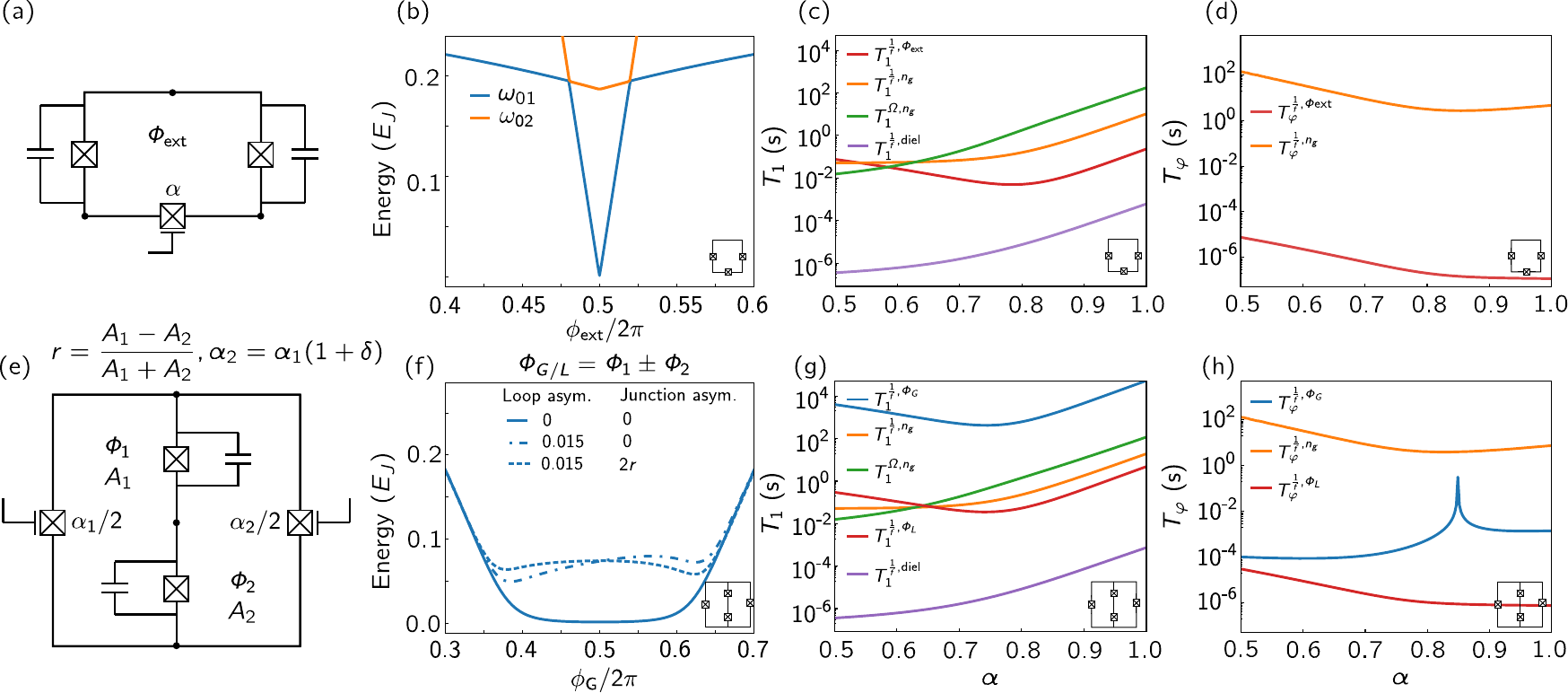}
    \caption{(a) Circuit layout of the single-loop DSFQ. (b) Dispersion of the qubit frequency with respect to the \rettelse{reduced} external flux through the loop, showing the linear dependence \rettelse{in the region $\phi_\text{ext}/2\pi = 0.47-0.53$.} (c-d) Relation between the relaxation/dephasing time ($T_1/T_\varphi$) and the barrier height controlled by $\alpha$ for the single loop DSFQ. (e) Circuit layout of the gradiometric DSFQ with two tunable junctions. The inconvenient placement of large capacitors in the loops can be worked around by using cross-over junctions. (f) Dispersion of the qubit frequency with respect to the \rettelse{redcued} external global flux\rettelse{, $\phi_G = \frac{2\pi \Phi_G}{\Phi_0}$}. We display the cases where the loop areas are identical (solid line), non-identical (dot-dashed line) and non-identical with compensating asymmetric junctions (dashed line). (g-h) Relation between the relaxation/dephasing time ($T_1/T_\varphi$) and the barrier height controlled by $\alpha$ for the gradiometric DSFQ. Note the insensitivity to noise in the global magnetic field and sensitivity to local magnetic field noise. The noise amplitudes in all figures are $A_\Phi = 10^{-6} \Phi_0/ \sqrt{\text{Hz}}$, $A_{n_g} = 10^{-4} e/ \sqrt{\text{Hz}}$ \cite{groszkowski_coherence_2018}, $B_{n_g} = 5.2 \times 10^{-9} e / \sqrt{\text{Hz}}$ \cite{yan_flux_2016} and $\tan\delta_\text{diel} = 2\cross 10^{-7}$ \cite{Nguyen2022blueprint}. The Josephson energy is $E_J = 10\, h\, \si{\giga\hertz}$ \rettelse{and external flux is $\phi_{ext} = 0.997 \pi$} where relevant.}
    \label{fig:fig2}
\end{figure*}

\subsection{Gradiometric DSFQ}

The qubit discussed above is designed to have a large relaxation time due to the exponential suppression of inter-well coupling. However, it is likely to have a poor dephasing coherence time because of the sensitivity of the energy difference of the two wells to flux noise. To improve the dephasing time, we propose a double-loop variation as in Fig.~\ref{fig:fig2}(e) which is designed to cancel out any fluctuations in the global flux. \rettelse{In the double-loop design, we picture the variable junctions as tunable nanowire junctions. Alternatively, these could be SQUIDs controlled by individual flux lines without defeating the purpose of the gradiometric setup. However, the additional flux loops will complicate the control of the qubit because there will be flux lines to each SQUID and one to control the global flux. The tunable Josephson junctions give an advantage with fewer flux control lines compared to using SQUIDs at the potential expense of reduced coherence due to semiconducting junctions.} To understand \rettelse{the double-loop cancellation} better, we consider the situation where half a flux quantum threads through each loop. This gives rise to two lowest-energy combinations of current flowing in the circuit; $\ket{\circlearrowleft \circlearrowright}, \ket{\circlearrowright \circlearrowleft}
$, where an arrow indicates the direction of the current in each loop. 
Thus, the two lowest energy states correspond to the situation where current flows in opposite directions, making them indifferent to variations in the external flux. Said differently, the magnetic dipole moment vanishes and the computational states are only affected by magnetic field gradients through the magnetic quadrupole moment as verified in Refs.\ \cite{schwarz_gradiometric_2013, kou_fluxonium-based_2017, gusenkova_operating_2022}. In Fig.~\ref{fig:fig2}(b,f), we show the dependence of the qubit splitting on the global flux for both single- and double-loop DSFQs.

For a symmetric situation where the areas of the two loops and the Josephson energies of two outer junctions are identical, the dependence of the global flux $\Phi_G$ (proportional to a global magnetic field) has zero slope when $\Phi_G$ \rettelse{is at half flux quantum} (see Fig.~\ref{fig:fig2}(f), blue solid line). In an experimental situation, the loop areas will be slightly different, leading to a sensitivity to the global magnetic field (blue dash-dotted line). However, by appropriately choosing the ratio of the tunable junctions, the dispersion with $\Phi_G$ can become exponentially flat again at the expense of splitting the degeneracy (blue dashed line). If the flux through the two non-identical loops is controlled by a single global field, and the tunable junctions can be tuned to be asymmetric, $\alpha_2 = (1 + \delta)\alpha_1$, then the sweet-spot simply shifts to
\begin{equation}\label{deltacondition}
  \delta=-1+\frac{1+r}{1-r}\cos\left(\frac{2\pi r}{1-r}\right) \approx 2r,\quad  r=\frac{A_1-A_2}{A_1+A_2}.
\end{equation}
where $r$ is a measure of the loop area asymmetry \rettelse{and assumed small}, see also Appendix \ref{app:gradiometric} where the condition on $\delta$ is derived. Here, it is also shown that the fluctuations in $\delta$ has very little effect on the energies near half a flux quantum as can also be seen by comparing the dashed ($\delta = 2r$) and dash-dotted blue line ($\delta = 0$). Fig.~\ref{fig:fig2}(f) summaries how the sensitivity to the external global magnetic field and how choosing the value of the Josephson energy of the second junction can make the spectrum practically insensitive to the global field. As detailed in Appendix \ref{app:gradiometric}, the slope and height of the curve is set by the area and junction asymmetry. While being insensitive to variations in the global magnetic field, the qubit frequency is still linearly sensitive to the local fluxes in the individual loops, see Fig.\ \ref{fig:fig2}(h) and discussion below.

\subsection{Decoherence times}

The decoherence of the DSFQ is estimated by calculating relaxation and dephasing rates for different noise sources. The relaxation time $ T_1= \left( \sum_\lambda \Gamma_1^\lambda\right)^{-1}$ is computed through the relaxation rates which are given by Fermi's Golden rule \cite{ithier_decoherence_2005,Nguyen2022blueprint,groszkowski_coherence_2018} 
\begin{align}
\Gamma_1^\lambda &= \frac{1}{\hbar^2} \left| \bra{1} \partial_\lambda H \ket{0}\right|^2 S_\lambda(\omega), \nonumber\\
\Gamma_1^{\text{diel}} &= \hbar \left| \bra{1} \phi \ket{0}\right|^2 S^{\text{diel}}(\omega)
\end{align}
where $\lambda$ is an external noise source and $S_\lambda(\omega)$ is the power spectral function for a given noise source. We consider $1/f$ and ohmic noise which were the limiting noise channels for flux and charge noise respectively for the CSFQ \cite{yan_flux_2016} in addition to dielectric loss, the limiting factor for fluxonium relaxation time \cite{nguyen_high-coherence_2019, Nguyen2022blueprint}. The associated spectral functions are
\begin{align*}
S_\lambda^{\frac{1}{f}}(\omega) &= \frac{2\pi A_\lambda^2 \text{Hz}}{|\omega|}, \quad S_\lambda^\Omega(\omega) = \frac{B_\lambda^2 \omega}{2\pi \times 1 \text{GHz}},
\end{align*}
\begin{equation*}
    S^\text{diel}(\omega) = \frac{\omega^2 \tan\delta_\text{diel}}{4E_C}\left[ \coth(\frac{\omega}{k_B T}) + 1 \right],
\end{equation*}
where $A_\lambda$ and $B_\lambda$ are noise amplitudes for $1/f$ and ohmic noise respectively, $\tan\delta_\text{diel} = 2\cross 10^{-7}$ is the loss tangent and $T =20$mK is the temperature \cite{Nguyen2022blueprint}. We use typical noise amplitudes $A_\Phi = 10^{-6} \Phi_0/ \sqrt{\text{Hz}}$ \cite{groszkowski_coherence_2018}, $A_{n_g} = 10^{-4} e/ \sqrt{\text{Hz}}$ \cite{groszkowski_coherence_2018} and $B_{n_g} = 5.2 \times 10^{-9} e / \sqrt{\text{Hz}}$ \cite{yan_flux_2016}.

In Fig. \ref{fig:fig2}(c, g), we display the computed relaxation times for the single loop and double loop (gradiometric) versions of the DSFQ. Both panels show exponentially enhanced $T_1$ in the protected regime ($\alpha = 1$) \rettelse{with $T_1 = 603$ $\mu$s in the single loop and $T_1 = 733$ $\mu$s in the gradiometric setup}, the limiting factor being dielectric loss. In the unprotected regime ($\alpha=0.5$), the relaxation time is reduced \rettelse{to $ T_1 = 0.35$ $\mu$s in the single loop and $T_1 = 0.35$ $\mu$s in the gradiometric equivalent to 3 orders of magnitude}.

We can compare the relaxation times to the dephasing times shown in Fig.\ \ref{fig:fig2}(d, h). The first order dephasing rates for $1/f$ noise are computed through \cite{groszkowski_coherence_2018}, 
\begin{align*}
\Gamma_\varphi^{\frac{1}{f},\lambda} = \sqrt{2A_\lambda \left(\partial_\lambda \omega_q\right)^2 \ln|\omega_\mathrm{ir}t|},
\end{align*}
where we have introduced an infrared cutoff and a characteristic time with the product $\omega_\mathrm{ir} t = 2\pi \times 10^{-6}$ as in Ref. \cite{groszkowski_coherence_2018}. The dephasing times shown in Fig.~\ref{fig:fig2}(d, h) are limiting the coherence time $\frac{1}{T_2} = \frac{1}{2T_1} + \frac{1}{T_\varphi}$ compared to the relaxation time due to the linear sensitivity to (local) flux noise in the $T_1$-protected regime. Conversely, in the unprotected regime, the coherence is limited by relaxation through dielectric loss, illustrating the trade-off between $T_1$-protection and dephasing due to flux noise \rettelse{is} general to flux qubits. Note that the sensitivity to global flux noise in Fig.~\ref{fig:fig2}(g, h) is reduced due to the gradiometric construction of the device. \rettelse{In the $T_1$ protected regime ($\alpha = 1$) the dephasing time is $T_\varphi = 0.12$ $\mu$s in the single loop and $T_\varphi = 0.74$ $\mu$s in the gradiometric setup. In the unprotected regime ($\alpha = 0.5$) the dephasing time is enhanced to $T_\varphi = 7.6$ $\mu$s in the single loop and $T_\varphi = 98$ $\mu$s in the gradiometric setup. The CSFQ has relaxations times reported in the range $T_1 = 20-60$ $\mu$s \cite{yan_flux_2016}. State of the art transmon qubit report relaxations times up to 
$T_1 = 0.5$ ms \cite{wang_towards_2022}.} 

In total, the DSFQ \rettelse{does not exceed the relaxation time of state of the art transmon qubits but} offers a platform with adjustable and strong noise bias and a tunable degree of $T_1$-protection, \rettelse{which can be used to study optimum strategies for gate operations on protected qubits}. While the noise bias\rettelse{, in principle,} opens up paths towards efficient noise biased error correcting codes, the linear sensitivity to (local) flux noise is a limiting factor. \rettelse{This could be suppressed by choosing a larger qubit splitting, creating a wider sweet spot at half flux quantum. However, we have chosen to focus on the $T_1$ protected regime here. We note that such compromise is relevant for other qubit proposals such as the heavy fluxonium and the bifluxon \cite{Earnest_heavy_fluxonium_2018, zhang_universal_2021, kalashnikov_bifluxon_2020}}.

\section{Qubit control}

To control the DSFQ, we leave the protected regime ($\alpha=1$) and lower the barrier between the two wells ($\alpha \simeq 0.5-0.7$). When the barrier is lowered, traditional techniques in microwave control such as DRAG and IQ-mixing can be used for the DSFQ \cite{Motzoi_DRAG_2009, krantz_quantum_2019}. As detailed in the sections below, the height of the barrier at the operating point and the rate at which it is lowered depends on whether single or two-qubit gates are performed. We continue in the following section by implementing an $\sigma_x$ gate numerically to illustrate how single qubit gates can be performed on qubits with variable-protection using single-tone driving.

\subsection{Variable-protection single qubit gates}

\begin{figure}
    \centering
    \includegraphics{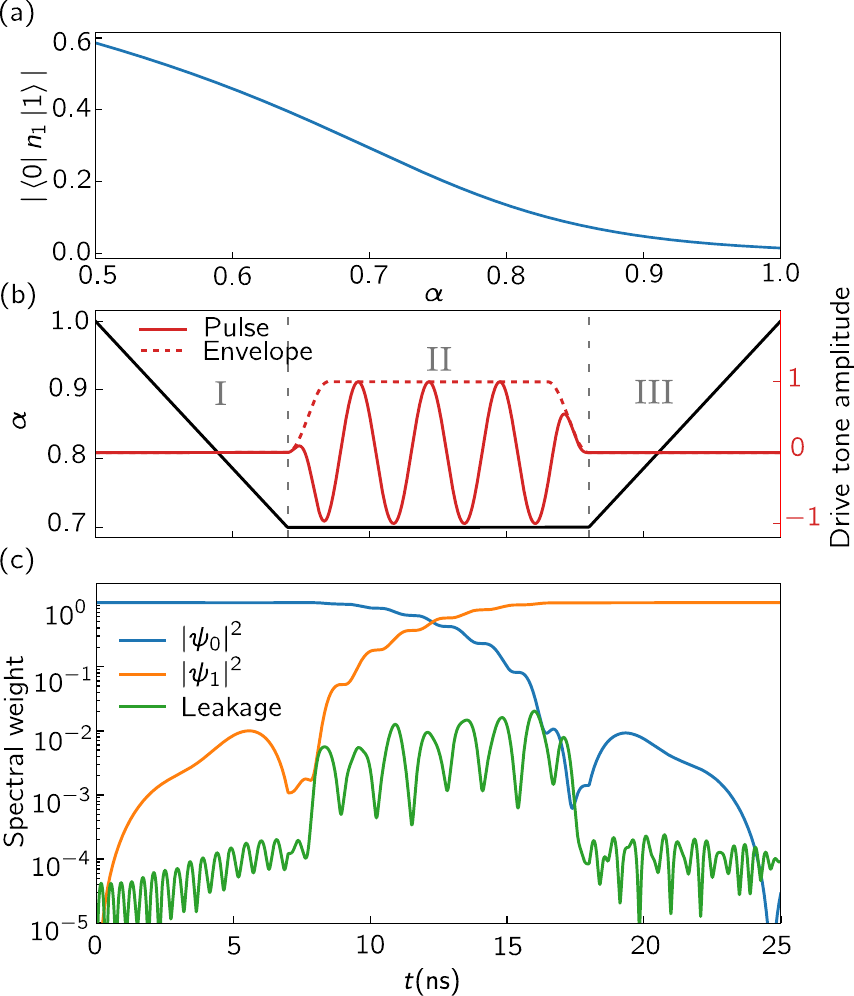}
    \caption{(a) The coupling of computational states through the charge operator as a function of $\alpha$, showing when transitions can be stimulated through a capacitively coupled drive-line. (b) The pulse profile for the $\sigma_x$ gate displaying the low-frequency $\alpha$ drive (black) and the high-frequency microwave drive (red). The envelope of the microwave pulse is $11\,\si{\nano\second}$ long with a $1.5\,\si{\nano\second}$ cosine ramp up/down. The drive frequency is slightly detuned from the qubit frequency $\omega_d = 0.979\, \omega_q$. (c) Numerical data \rettelse{from non-dissipative simulations} showing the time history of the spectral weights during the low-leakage, high fidelity $\sigma_x$ gate. In all panels, the scale of the Josephson energy is $E_J=10\,h\,\si{\giga\hertz}$ and $E_J/E_C = 100$ with the flux bias set to $\phi_\text{ext} = 0.995\pi$.}
    \label{fig:singlequbitgate}
\end{figure}

Our proof-of-concept $\sigma_x$-gate has three steps as illustrated in Fig.\ \ref{fig:singlequbitgate}: 
\begin{enumerate}[I.]
    \item Lower the barrier adiabatically,  $\alpha=1\to0.7$.
    \item Apply an appropriate microwave pulse to the qubit.
    \item Raise the barrier adiabatically, $\alpha=0.7\to1$.
\end{enumerate}
This control sequence is illustrated in Fig.\ \ref{fig:singlequbitgate} where the lowering and raising of the barrier takes $7\,\si{\nano\second}$ and the microwave drive takes $11\si{\nano\second}$ (including $1.5\,\si{\nano\second}$ ramp up/down), totalling a gate time of $25\, \si{\nano\second}$. The microwave drive line is coupled to one of the nodes of the qubit through a small capacitance $C_d \ll C$, giving rise to the Hamiltonian term $H_d =  \frac{C_d}{C+C_d}V_d(t) n_1$ \cite{krantz_quantum_2019}. \rettelse{As the barrier is lowered, the quantum states changes significantly and a small subspace of states is insufficient to describe the evolution due to $H(t)=H_C+H_J(t)$. We therefore perform simulation in a relatively large Hilbert space with $625$ states (in the charge basis with cutoff $n_\mathrm{cutoff}=12$ for both the $\phi$- and $\theta$-mode) and numerically evaluate $\exp(-i\Delta t\, H(t))$ at each time step to perform the time-evolution (857 timesteps/nanosecond). When the drive is turned on at fixed $\alpha$, we instead numerically integrate the time-dependent Schrödinger equation using the same Hilbert space dimension. At each time-step, we numerically diagonalize the Hamiltonian and compute the overlap with the instantaneous qubit states to produce Fig.~ \ref{fig:app-single-gate}(c).}

\rettelse{In our single qubit gate scheme,} we choose to lower the barrier only partially ($\alpha=0.7$) to limit the time spent adiabatically adjusting $\alpha$ and to avoid small, unwanted interactions with neighboring qubits which arise when the barrier is completely lowered, see also Appendix \ref{app:two-gate}. \rettelse{The qubit frequency is changed from $\omega_q(\alpha = 1) = 0.25 \,h\, \si{\giga\hertz}$ to $\omega_q(\alpha = 0.7) = 0.39 \,h\, \si{\giga\hertz}$, where the Josephson energy is $E_J = 10 \,h\, \si{\giga\hertz}$ and $E_J/E_C = 100$ with the flux bias set to $\phi_{ext} = 0.995 \pi$. At the operating point ($\alpha =0.7$) the relaxation time is reduced to $1.6\, \si{\micro\second}$.} The speed at which the barrier is lowered is adiabatic with respect to the energy gap between the computational states and the non-computational states such that the adiabatic time is set by the desired leakage bound. The $7\,\si{\nano\second}$ lowering time results in a very small ($\sim10^{-4}$) leakage but does admit for a small ($\sim 10^{-3}$) probability to transition from one logical state to the other. This small effect makes it necessary to slightly adapt the microwave pulse to achieve high fidelity. \rettelse{One possibility is to marginally reduce the drive amplitude, but the qubit frequency is also shifted due to the AC-Stark effect. We therefore instead adapt the pulse} by a minor frequency shift of the drive, $\omega_d = 0.979\,\omega_q$\rettelse{, to account for both of these contributions}. \rettelse{The limit to the fidelity imposed by coherent errors (leakage) during the $\sigma_x$ gate} is $99.98\,\%$ while the gate time is \rettelse{$T_g = 25\, \si{\nano\second}$}. \rettelse{The single qubit gate fidelity is limited by decay from the shorter relaxation time at the operating point. We estimate the $T_1$ limited fidelity via $F\approx \exp[-\int_0^{T_g} \dd t\, \Gamma_1(t)]$, where $\Gamma_1(t)$ is the sum of (instantaneous) decay rates. The resulting $T_1$-limited fidelity is $99.1\%$ for the single qubit X-gate.} \rettelse{While the gate is limited by decay in this device, the coherence limited gate fidelity} is comparable to state-of-the-art single qubit gates on unprotected qubits such as the transmon \cite{krinner_realizing_2022} and potentially faster than alternative gates on $T_1$-protected qubits \cite{abdelhafez_universal_2020}. The latter makes use of non-computational states, multi-tone driving and an optimal control algorithm to optimize gate performance. The comparatively simple \rettelse{variable-protection} gate shows the benefits of tuning in and out of protection, and that the access to fast, single tone pulse control outweigh the additional overhead from the adiabatic control of the level of protection. In Appendix \ref{app:single-gate}, we exemplify using standard IQ-mixing how also $\sigma_y$ and $(\sigma_x - \sigma_y)/\sqrt{2}$ gates can be implemented with similar fidelity as the $\sigma_x$ gate. Combined with virtual $\sigma_z$ gates, we have thus demonstrated a compelling scheme for realizing universal single-qubit control. It is natural to improve upon this proof-of-principle design using more advanced $\alpha$-profiles combined with microwave pulse shaping techniques such as DRAG \cite{Motzoi_DRAG_2009} \rettelse{in order to reduce the time spent at low coherence for smaller $\alpha$}. Alternatively, sudden gates or gates where the flux bias is also controlled may be explored with inspiration from Ref.\ \cite{zhang_universal_2021}. Ref.\ \cite{zhang_universal_2021} also shows how multi-tone driving can initialize low-frequency qubits where the qubit frequency is subthermal. Alternatively, our flexible design also allows for thermal initialization in the unprotected regime.

\subsection{Variable-protection two-qubit gates}

\begin{figure*}[t]
    \centering
    \includegraphics{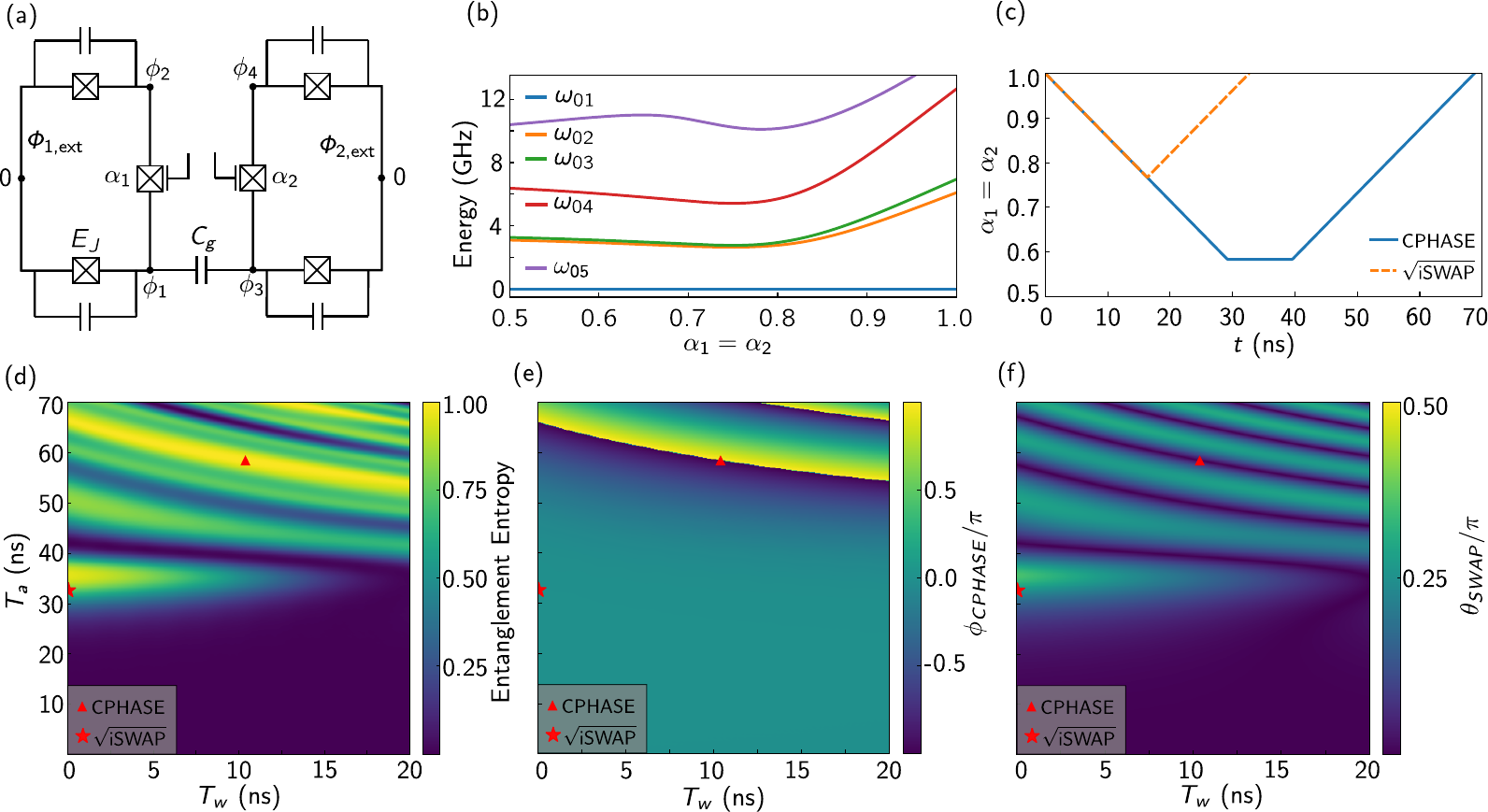}
    \caption{Two qubit setup and gate characteristics \rettelse{from non-dissipative simulations}. (a) Schematic of two capacitively coupled DSFQs with substantial coupling capacitance $C_g = 0.3\, C$. (b) The five lowest energy states shown as the two barriers are lowered simultaneously by decresing $\alpha_1=\alpha_2$. (c) The $\alpha_1 = \alpha_2$ profile as a function of time for the CPHASE gate. (d) The entanglement entropy of the final two-qubit gate as a function of the waiting time $T_w$ and the total adiabatic control time $T_a$. 
    The red markers in this and subsequent panels show the optimal $\sqrt{\text{iSWAP}}$ (star) and CPHASE (triangle) gates which have respective fidelities \rettelse{limited by coherent gate errors} and gate times of $\mathcal{F}_{\sqrt{\text{iSWAP}}}=99.96\,\%$, $T_{\sqrt{\text{iSWAP}}} = 32.66\,\si{\nano\second}$ and $\mathcal{F}_{\text{CPHASE}}=99.95\,\%$, $T_\text{CPHASE} = 68.76\,\si{\nano\second}$. \rettelse{The estimated $T_1$-limited fidelities are $F_{\sqrt{\text{iSWAP}}} = 99.7\%$ and $F_{\text{CPHASE}} = 91.4\%$}. (e-f) The resulting phase and swap parameters $\phi_\text{CPHASE}$ and $\theta_\text{SWAP}$ of the final two-qubit gate as a function of the waiting time $T_w$ and the total adiabatic control time $T_a$. \rettelse{The flux bias is set to $\phi_\mathrm{ext} = 0.99\pi$.}
    }
    \label{fig:twoqubitgate}
\end{figure*}

An advantage of qubits with variable protection is that they can act as their own tunable couplers with exponential on/off ratio. In the protected idling regime, the qubit-qubit coupling vanishes due to the exponentially small wave-function overlap, see also Appendix \ref{app:two-gate}. As a result of the exponentially suppressed coupling between the computational states in the idling regime, a capacitive qubit-qubit coupling,
\begin{equation}\label{eq:QQcoupling}
    H_{\mathrm{Q-Q}} = 4E_C \frac{C_g}{C+C_g} n_1 n_3,
\end{equation}
may be relatively strong $C_g\simeq 0.3\, C$ compared to e.g.\ transmon qubits, see Appendix \ref{app:two-gate} for a derivation of Eq.\ \ref{eq:QQcoupling}. We can thus implement two-qubit gates that rely solely on the simultaneous lowering of both barriers of two capacitively coupled DSFQs. 

Our implementation of two-qubit gates has three steps:
\begin{enumerate}[I.]
    \item Lower both barriers simultaneously in a time $T_a/2$,
    $\alpha_1=\alpha_2=1\to\alpha_\mathrm{min}$.
    \item Wait for a time $T_w$.
    \item Raise the barriers simultaneously  in a time $T_a/2$,
    $\alpha_1=\alpha_2=\alpha_\mathrm{min}\to1$.
\end{enumerate}
The total gate time thus becomes the sum of the waiting time and the adiabatic control time, $T_{2Q}=T_a+T_w$. 

When the barriers are lowered, the qubits can exchange excitations through the capacitive coupling element resulting in an effective $\sigma_x^{(1)} \sigma_x^{(2)}+\sigma_y^{(1)} \sigma_Y^{(2)}$ interaction. 
Crucially, the adiabatic control time can be adjusted such that there occurs a transition between the states $\ket{01}$ and $\ket{10}$ due to their small energy difference and not between other computational states whose energy difference is large compared to the adiabatic time. 
As shown in Fig.\ \ref{fig:twoqubitgate}(b), \rettelse{an} avoided crossing occurs near $\alpha = 0.75$. \rettelse{On the other side of this avoided crossing, when $\alpha$ is further decreased, the coupling dramatically increases. See also Fig.\ \ref{fig:appendix_coupling}, where the $\sigma_z^{(1)} \sigma_z^{(2)}$-interaction strength is shown.} \rettelse{The avoided crossing shown in Fig.\ \ref{fig:twoqubitgate}(b) is a generic feature of the coupled spectrum as long as the qubit frequencies of the two interacting qubits are similar at $\alpha = 1$.}

To exclude transitions between the other computational states and transitions out of the computational subspace, the speed at which $\alpha$ is lowered should be slower compared to the single qubit gate. As a concrete example, we consider lowering the barriers with a constant speed, meaning that the adiabatic time is proportional to the minimum value $\alpha_\mathrm{min} = 1-\frac{T_a/2}{2\cdot 35\,\si{\nano\second}}$. Thus, the barrier can be completely lowered in $35\, \si{\nano\second}$ which is three times slower than for the lowering rate used for the single qubit gate. Adiabatic lowering/raising times $T_a/2$ less than $35\, \si{\nano\second}$ results in only partly lowering the barrier due to the constant lowering/raising speed, see also Fig.\ \ref{fig:twoqubitgate}(c). 

In addition to the $\sigma_x^{(1)} \sigma_x^{(2)} + \sigma_y^{(1)} \sigma_y^{(2)}$ interaction, the energies of the coupled system shifts relative to the bare energies due to an effective $\sigma_z^{(1)} \sigma_z^{(2)}$ interaction, see also Appendix \ref{app:two-gate}. \rettelse{Below we simulate the two-qubit gate shown in Fig.\ \ref{fig:twoqubitgate} and discuss the types of gates achieved. The two-qubit unitaries can be modelled by a two-qubit interacting system of the following form}
\begin{align}\label{eq:effmodel}
    H_{\mathrm{eff}} &= -\frac{\omega_1}{2}\sigma_z^{(1)}
    -\frac{\omega_2}{2}\sigma_z^{(2)}\\
    &+\frac{g_{xy}}{2}\left(
    \sigma_x^{(1)}\sigma_x^{(2)} 
    + \sigma_y^{(1)}\sigma_y^{(2)} 
    \right)
    + \frac{g_z}{2}\sigma_z^{(1)}\sigma_z^{(2)},\nonumber
\end{align}
\rettelse{where the $\sigma_{x,y,z}^{(i)}$'s are Pauli matrices acting in the logical subspace of qubit $i$, $\omega_i$ describe the qubit frequencies, and the swap coupling $g_{xy}$ and $\sigma_z^{(1)}\sigma_z^{(2)}$ coupling $g_z$ are all $\alpha$-dependent. This model Hamiltonian} gives rise to the so-called fSim-gates which interpolate between the iSWAP- and CPHASE-gate \cite{krantz_quantum_2019, google_ai_quantum_demonstrating_2020},
\begin{equation}
    U_\text{fSim} = 
        \begin{pmatrix}
            1 & 0 & 0 & 0 \\
            0 & \cos(\theta_\text{SWAP}) & -i\sin(\theta_\text{SWAP}) & 0 \\
            0 & -i\sin(\theta_\text{SWAP}) & \cos(\theta_\text{SWAP}) & 0 \\
            0 & 0 & 0 & e^{-i \phi_\text{CPHASE}}
        \end{pmatrix}
\end{equation}
\rettelse{which is precisely what we see in the simulation of the full model.}
By timing the adiabatic control time and the waiting time to match $\ket{01} \Longleftrightarrow \ket{10}$ swap oscillations and the rotating $\sigma_z^{(1)}\sigma_z^{(2)}$-phase, different gates in the fSim-space can be targeted as shown in Fig.\ \ref{fig:twoqubitgate}(d-f). Here, we sweep over the adiabatic and waiting times, $T_a$ and $T_w$, and in panel (d), we display the entanglement entropy which is normalized to unity for maximally entangling gates \cite{Zanardi_2000_entanglingpower}. The only maximally entangling gates in the fSim-space are CPHASE and iSWAP. In panels (e) and (f), we decompose the resulting unitary into the fSim-parameters; the phase angle $\phi_\text{CPHASE}$ and the swap angle $\theta_\text{SWAP}$. The red markers show two example gates in the fSim-space; the CPHASE and $\sqrt{\text{iSWAP}}$ gates. \rettelse{The fidelity limited by coherent errors (lekage) is } well beyond $99.9\%$ (up to single qubit $\sigma_z$-gates) and can be performed in about 69 ns and 33 ns respectively. \rettelse{Again, the two-qubit gates are limited by decay, with estimated $T_1$-limited fidelities of $F_{\sqrt{\text{iSWAP}}} = 99.7\%$ and $F_{\text{CPHASE}} = 91.4\%$. The fidelity of the CPHASE gate is severely impacted by the low qubit coherence near $\alpha=0.6$ where $T_1=0.6\,\si{\micro\second}$ but the $\sqrt{\text{iSWAP}}$ gate is a promising high fidelity alternative.} \rettelse{The iSWAP gate cannot be implemented to high fidelity} \rettelse{as it requires both fine-tuning of energies to achieve a full swap of excitations and zero (mod $2\pi$) $\sigma_z^{(1)} \sigma_z^{(2)}$-phase}. \rettelse{The combined requirement is challenging to tune with our parameters, so we instead propose to} simply apply two $\sqrt{\text{iSWAP}}$ gates successively. \rettelse{The $\sqrt{\text{iSWAP}}$ gate is comparatively easy to perform as a partial swap of excitations happens before any significant $\sigma_z^{(1)}\sigma
_z^{(2)}$ phase is accrued. Finally, The CPHASE gate depends to an intermediate degree on the Hamiltonian parameters as it does not require a transfer of excitations. Our testing finds that appropriate times $T_a$ and $T_w$ can be chosen for a range of parameters to yield a CPHASE gate.}

\rettelse{As mentioned, the wave functions change substantially as the barriers are lowered and complicates the simulation of the qubit interactions. In order to faithfully simulate the time-evolution, we numerically diagonalize the charge-basis Hamiltonian ($n_\mathrm{cutoff} = 9$) at each $\alpha$ and keep the 24 lowest states. Since the diagonalizing unitary, $V:\, V^\dagger H V = \mathrm{diag}(E_1, E_2, \ldots)$, is time-dependent, the Schrödinger equation acquires an additional term, $- i V^\dagger \partial_t V$. Finally, using the combined Hamiltonian $H = H_1 + H_2 + H_\mathrm{Q-Q}$ (Eqs.\ (\ref{Hpot}), (\ref{Hkin}) and \eqref{eq:QQcoupling}), the time-evolution operator of the lowest 24 states is evolved by $\exp\left[-i(V^\dagger H V - i V^\dagger \partial_t V)\Delta t\right]$ at each timestep $\Delta t$ (286 timesteps/nanosecond).}

Despite relying only on adiabatic control, the two-qubit gates presented here are competitive compared to state-of-the-art two-qubit gates for both single- and double-well qubits \cite{zhang_universal_2021, ficheux_fast_2021, gyenis_experimental_2021}. Further advantages include the exponential on/off coupling ratio, that only the computational states are used and the possibility of being able to produce different gates in the fSim-space. Further developments, for example controlling $\alpha_1$ and $\alpha_2$ individually as well as the fluxes, will likely provide more control over what fSim-gates can be reached and reduce the overall gate time or increase fidelities using optimized strategies. Additionally, recent work suggests to also use the DSFQ as a transmon-transmon coupler (called the ``double transmon coupler''), which illustrates the exciting flexibility of the device \cite{Goto2022doubletransmoncoupler}.

\subsection{Readout}

\begin{figure}
    \centering
    \includegraphics{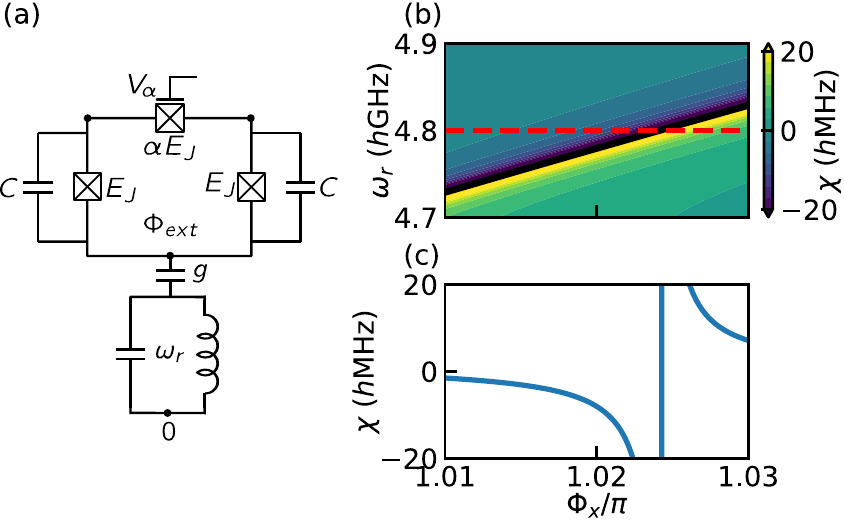}
    \caption{
    (a) 
    The qubit coupled to a readout resonator. The qubit induces a state-dependent shift of the frequency of the resonator, which can be measured using standard techniques.
    (b)
    Dispersive shift as a function of external flux. By adjusting the flux away from half flux bias, a resonance between one of the computational states become and a higher energy states comes close to the frequency of the readout resonator. The resonator shift is increased resulting in a stronger readout signal. A smaller shift is preferable in the context of error suppression where it reduces the sensitivity to photon-shot noise.}
    \label{fig:fig5}
\end{figure}

Readout of the DSFQ device can be performed using conventional dispersive readout techniques \cite{krantz_quantum_2019}. \rettelse{However, rather than reading out via the $\phi$-mode, similar to fluxonium qubits, we instead propose to readout via the $\theta$-mode. By} coupling the qubit capacitively to a readout resonator through the $\theta$ degree of freedom, as shown in Fig. \ref{fig:fig5}(a)\rettelse{, we can achieve substantial dispersive shifts while remaining in the protected qubit regime to suppress (Purcell enhanced) relaxation}. As we detail below, the plasmon frequency of the $\theta$-mode depends on which well the $\phi$-mode is localized in. Further, the difference in plasma frequencies for the two wells are tuned by the external magnetic flux. In this way, we can use the external flux to control the state dependent shift of the readout resonator as shown in Fig.\ \ref{fig:fig5}(b-c).

We start by considering the Hamiltonian of the combined system which can be written as \cite{krantz_quantum_2019}
\begin{equation}
    H = H_{\mathrm{sys}} + g(a + a^\dag)n_\theta+\omega_r a^\dag a,
    \label{eq:readout_hamil_raw}
\end{equation}
where $H_{\mathrm{sys}}$ is the qubit Hamiltonian, $a(a^\dag)$ is the resonator annihilation(creation) operator, $\omega_r$ is the bare resonator frequency and $g$ is the coupling strength between resonator and qubit. 
In the dispersive regime, the resonator frequency is effectively shifted by the state of the qubit. 
This can be seen by performing a Schrieffer-Wolff transformation \cite{blais_circuit_2021,zhu_circuit_2013} to second order,
\begin{equation}
    H_{\mathrm{eff}} = H_{\mathrm{sys}} + \omega_r a^\dag a - \left(\frac{\chi}{2} a^\dag a + \delta \right)\sigma_z,
\end{equation}
where $\chi$ is the qubit state dependent resonator shift, $\delta$ is a small shift of the qubit frequency and $\sigma_z=\ket{0}\bra{0}-\ket{1}\bra{1}$ is the qubit Pauli $Z$ operator. 
To correctly estimate the dispersive resonator shift it is important to account for higher levels outside of the computational subspace. 
Carrying out the perturbation calculation, we find the dispersive shift as $\chi = \sum_j\chi_{1j}-\chi_{0j}$, where
\begin{equation}
    \chi_{ij} = g^2\left|\bra{i}n_\theta\ket{j}\right|^2\left(\frac{1}{E_i-E_j-\omega_r}+\frac{1}{E_i-E_j+\omega_r}\right).
\end{equation}
Figure \ref{fig:fig5}(b, c) shows the resonator shift as a function of the externally applied magnetic flux.
For these simulations, we have used a bare resonator frequency of $\omega_r=4.8\,h\,\si{\giga\hertz}$ and coupling strength of $g = 25\,h\,\si{\mega\hertz}$.

To \rettelse{explain the working principle of the readout, we briefly adopt a simple, minimal model of the DSFQ. In this model, we assume that we are away from the sweet-spot at exactly half flux quantum and write an} effective potential for the $\theta$-degree of freedom by freezing the $\phi$-degree of freedom to one of the two minima at $\phi_\pm = (\pm \pi - \delta\phi_{ext})/3$ for $\alpha=1$ and thus \rettelse{momentarily} neglect tunneling between the two wells, 
\begin{equation}\label{eq:readoutpot}
    V_\pm = \mp E_J\frac{\delta\phi_{ext}}{2\sqrt{3}} - E_J\left(1\pm\frac{\delta\phi_{ext}}{\sqrt{3}}\right) \cos(\theta),
\end{equation}
where $\phi_{ext} = \pi + \delta\phi$ and $\delta\phi\ll1$.
In this picture, each minima corresponds to one of the computational states. \rettelse{Close to} half flux bias ($\delta\phi_{ext} \approx 0$), $V_\pm$ are \rettelse{nearly} identical and the readout resonator cannot discriminate between the computational states \rettelse{as the matrix elements $|\bra{\pm}n_\theta\ket{j}|$ are approximately the same for the two qubit states $\ket{\pm}$}. By increasing the offset from the flux frustration point, the two terms in Eq.\ \eqref{eq:readoutpot} lead to differences between the two wells that can result in a large dispersive shift \rettelse{if the readout resonator is close in frequency to the plasma frequency of the $\theta$-mode in one of the wells}. The first term \rettelse{in Eq.\ \eqref{eq:readoutpot}} contains the simple energy splitting between the two wells due to the external flux which does not change the plasmon frequency. The second term in Eq.\ \eqref{eq:readoutpot} shows that the plasmon frequency of the $\theta$-mode in each well $\omega_\theta^\pm = \sqrt{8 \widetilde E_C \widetilde E_J^\pm}$, where $\widetilde E_C$ and $ \widetilde E_J^\pm$ are the effective charging and Josephson energies of the $\theta$-mode \cite{krantz_quantum_2019}, also depends on the offset from half flux bias. In this way, we may tune the plasmon frequency in one of the wells close to the readout resonator frequency and thereby achieve a large dispersive shift, see Fig.\ \ref{fig:fig5}(b-c). \rettelse{We may now consider what happens at exactly half flux quantum where the small tunneling between the wells results in wave functions that are even/odd in $\phi$. In this situation, different selection rules for the even/odd computational states dictate what matrix elements can be nonzero and will generally result in a nonzero dispersive shift. However, as the resonator frequency can be far off the frequency of the contributing transitions, the dispersive shift remains small.}

There are several advantages to performing readout in \rettelse{the} proposed scheme: \rettelse{Suppression of the} dispersive shift \rettelse{controlled by} the external flux grants us insensitivity to dephasing through photon shot noise \cite{krantz_quantum_2019}. By coupling the readout resonator to the $\theta$-mode of the qubit, we \rettelse{also} obtain protection against Purcell decay\rettelse{: T}he matrix element $\bra{0}n_\theta\ket{1}$ \rettelse{(or in the notation surrounding Eq.\ \eqref{eq:readoutpot}), $\bra{+}n_\theta\ket{-}$) is zero since the computational states are both in the even $\theta$-mode ground state in their respective wells. Via this mechanism, the qubit is protected from the Purcell effect} due to the symmetries of the wave functions. \rettelse{There are no additional Purcell effect due to $n_\phi$ as the readout resonator remains decoupled from this mode. In total, the dominant source of error during readout is the direct tunneling between the qubit states. The $T_1$-times computed in Sec.\ \ref{sec:theory} depends weakly on the external flux and for readout at $\phi_\mathrm{ext} = 1.023\pi$ we find $T_1 = 519\, \si{\micro\second}$. For a readout integration time around $1\, \si{\micro\second}$, the $T_1$-limited readout fidelity is $F = 99.8\%$.}

\section{Conclusions and discussions}

In this paper, we have shown how gates and readout can be performed on a new flux qubit variation with a variable level of $T_1$-protection, the DSFQ. By adiabatically reducing the height of the barrier, the otherwise insensitive qubit can be made sensitive to a microwave drive. Our implementation of this \rettelse{variable-protection} gate scheme shows that fast, high fidelity single qubits gates can be performed without involving lossy non-computational states. We achieve single qubit gate\rettelse{s with coherence limited} fidelities at $99.98\%$ in $25\, \si{\nano\second}$, making it competitive with established gate schemes for both protected and unprotected qubits. \rettelse{However, non-optimized gates suffer from $T_1$ decay during the lowering of the barrier and results in a $T_1$-limited gate fidelity of $99.1\%$.} Likewise, we show that by lowering the barriers of two capacitively coupled DSFQs, that high fidelity two qubit gates in the fSim-space can be performed. Specifically, we find CPHASE and $\sqrt{\text{iSWAP}}$ gates \rettelse{with a coherence limited fidelity} above $99.9\%$ in $69\, \si{\nano\second}$ and $33 \, \si{\nano\second}$ respectively without residual $ZZ$-interactions. \rettelse{Again, the two-qubit gates are limited by relaxation and the $T_1$-limited fidelities are $F_{\text{CPHASE}} = 91.4\%$ and $F_{\sqrt{\text{iSWAP}}} = 99.7\%$ respectively} The fidelities and gate times can be further improved by using optimized protocols.

We have further shown that readout can be performed efficiently in the $T_1$-protected regime by adjusting the external flux bias away from the flux frustration point. Near half a flux bias, the dispersive shift is not only reduced due the the qubit-resonator detuning, but also due to the approximate symmetry between the two wells. With the order-of-magnitude variations in dispersive shift and separated double-wells, the DSFQ is robust againt noise channels arising from the coupling to the resonator.

In addition to the $T_1$-protection, we have also proposed a gradiometric double-loop variation of the DSFQ which is exponentially insensitive to global flux noise while remaining linearly sensitive to local flux noise. We show that area variability of the loops can be compensated for by making the tunable junction slightly asymmetric without being sensitive to the noise in the tunable junctions.

In total, the DSFQ presents an experimentally available platform for studying qubits with a variable level of $T_1$-protection, where gates can be performed without involving non-computational states. This contribution may help pave the way for achieving fast, high fidelity gates on protected qubits using this novel gate implementation.

\section{Acknowledgments}

We acknowledge helpful discussions from Andr{\'a}s Gyenis and Jonas Vinther and are grateful to Terry P.\ Orlando for comments on the manuscript. This research was supported by the Danish National Research Foundation, the Danish Council for Independent Research $|$ Natural Sciences. This project has received funding from the European Research Council (ERC) under the European Union’s Horizon 2020 research and innovation programme under Grant Agreement No. 856526. We acknowledge support from the Deutsche Forschungsgemeinschaft (DFG) – project grant 277101999 – within the CRC network TR 183 (subproject C03). MK gratefully acknowledges support for this research in part by the U.S. Army Research Office Grant No. W911NF-22-1-0042 and in part by the Villum Foundation (grant 37467) through a Villum Young Investigator grant.

\appendix
\begin{figure*}
    \centering
    \includegraphics{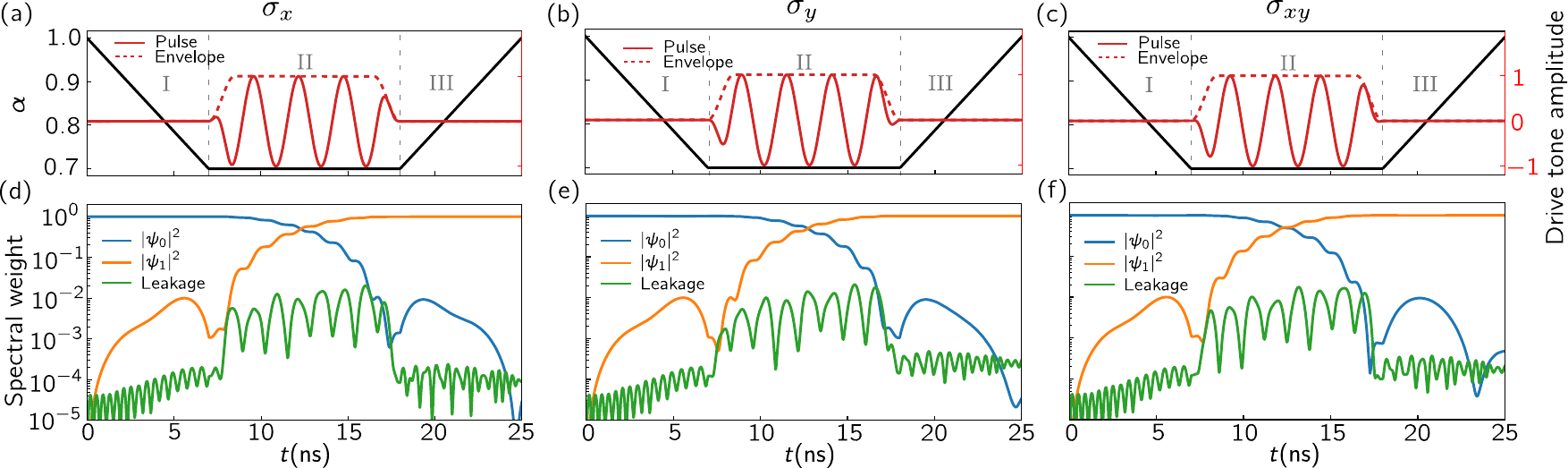}
    \caption{(a)-(c) Pulse sequence for the $\sigma_x$, $\sigma_y$ and $\sigma_{xy} = (\sigma_x-\sigma_y)/\sqrt{2}$. Parameters for the pulse envelope and the $\alpha$-profile is identical to those in Fig.\ \ref{fig:singlequbitgate}. The phase offset and drive frequency are in the three cases: (a) $\phi_\text{offset} = 0\pi, \omega_d = 0.979 \omega_q$, (b) $\phi_\text{offset} = 0.5\pi, \omega_d = 0.979 \omega_q$ and (c) $\phi_\text{offset} = 0.26\pi, \omega_d = 0.977 \omega_q$. (d)-(f) Corresponding evolution of the states during the gate operation. The fidelities in the three panels are (d) $F_x = 99.98\%$, (e) $F_y = 99.98\%$ and (f) $F_{xy} = 99.93\%$.}
    \label{fig:app-single-gate}
\end{figure*}
\begin{figure}
    \centering
    \includegraphics{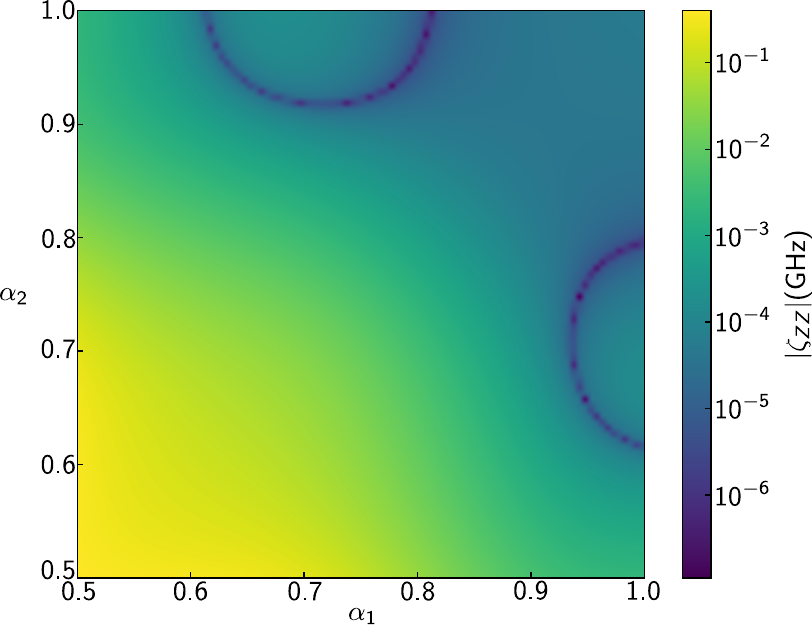}
    \caption{Plot of the $ZZ$-interaction strength due to the capacitive coupling. When one or none of the barriers are lowered, the interaction strength is suppressed to the $10 \, \si{\kilo\hertz}$ level.}
    \label{fig:appendix_coupling}
\end{figure}

\section{Gradiometric DSFQ}\label{app:gradiometric}

To better understand the dependence on the global flux, we look at the potential energy for the double-loop qubit
\begin{equation}\label{HpotDoubleLoop}
\begin{aligned}
  H_J =& -E_J \cos(\phi_1)-E_J \cos(\phi_2)+H_{J\alpha},\\
  H_{J\alpha}=&-\frac{E_J}{2}\left[\alpha_1\cos(\phiexta+2\phi)+\alpha_2\cos(\phiextb+2\phi)\right].
\end{aligned}
\end{equation}
Here the flux-induced phases are given by
\begin{equation}\label{varphiextsdef}
  \phi_{\mathrm{ext}1,2} = \frac{\pm 2\pi A_{1,2} B_{1,2}}{\Phi_0},
\end{equation}
where $A_{1,2}$ are the areas of the two loops and $B_{1,2}=(B\pm b)/2$ are the field through them. 

In the symmetric case when $A_1B_1=A_2B_2$, $\alpha_1=\alpha_2$ the flux-dependent term becomes
\begin{equation}\label{HJalpha}
    H_{J\alpha}=-\frac{E_J}{2}\alpha_1\cos(2\phi)\cos(\phiexta),
\end{equation}
and we see that the potential maintains the symmetry with two degenerate minima for all values of the global field $B$. However, it is not realistic to assume that the two areas can fabricated to be identical. Therefore, we consider the situation where they differ by some (small) amount. To study this case, we write $H_{J\alpha}$ as
\begin{equation}\label{HJalpha2}
    H_{J\alpha}=-\frac{E_J}{2}\left[V_c\cos(2\phi)-V_s\sin(2\phi)\right],
\end{equation}
where
\begin{subequations}\label{VcVsdef}
\begin{align}
  V_c &=  \alpha_1\cos(\phiexta)+\alpha_2\cos(\phiextb),\\
  V_s &=   \alpha_1\sin(\phiexta)+\alpha_2\sin(\phiextb).
\end{align}
\end{subequations}
The splitting of the degeneracy of the minima of $V_c$ is controlled by the second term $V_s$. One could, in principle, choose a set parameters  $(\alpha_1,\alpha_1, B_1,B_2)$ such that $V_s=0$ and regain the degenerate double-well potential. However, the degeneracy is lifted linearly in both the global external field $B$ and the tuning of the Josephson junctions, and the situation is therefore worse than before. Instead, we search for a point where the qubit is split by the different well depths, but with at least quadratic protection against deviations from the mentioned set of parameters. If both junctions in the outer SQUID-loop are tunable junctions, we have to minimize with respect to both which gives the condition $\sin(\phiexta)=\sin(\phiextb)$ at the operating point. Consequently, the condition for the junctions when minimizing with respect the global field $B$ is
\begin{equation}\label{aalphacondition}
  A_1\alpha_1=A_2\alpha_2.
\end{equation}
If the tunable junctions are parameterized as $\alpha_2=(1+\delta)\alpha_1$, the condition obtaining the sweet spot where the splitting is quadratic or better in $\delta$ and $B$ is
\begin{subequations}\label{sweetsplotdeltaB}
\begin{align}\label{eq:dvddelta}
   \frac{\partial V_s}{\partial \delta}\ = 0 &\to\ \sin(\phiextb) =0,\\
   \frac{\partial V_s}{\partial B} = 0 \ &\to\ \cos(\phiexta)=\frac{(1+\delta)A_2}{A_1}\cos(\phiextb).
\end{align}
\end{subequations}
Note that the condition in Eq. \eqref{eq:dvddelta} results in a $V_s$ which is insensitive to $\delta$ for all $\delta$.
If the flux through the two loops is controlled by a single global field (i.e., $b=0$), the two equations above can be combined to give the following condition on $\delta$,
\begin{equation}\label{deltacondition}
  \delta=-1+\frac{1+r}{1-r}\cos\left(\frac{2\pi r}{1-r}\right) \approx 2r,\quad  r=\frac{A_1-A_2}{A_1+A_2},
\end{equation}
for small $r$. 

\section{IQ-mixing}\label{app:single-gate}
We show that our single qubit gate scheme is compatible with IQ-mixing in Fig.\ \ref{fig:app-single-gate}. The pulses are parametrized by $\varepsilon(t) \cos(\omega_d t +\phi_\text{offset})$, where $\varepsilon(t)$ is the envelope with cosine ramp up/down and $\phi_\text{offset}$ is the phase offset that determines the $I$ and $Q$ components. We display three flip gates $\sigma_x$ (also found in Fig.\ \ref{fig:singlequbitgate}), $ \sigma_y$ and $\sigma_{xy} = (\sigma_x - \sigma_y)/\sqrt{2}$ with similar fidelities $>99.9\%$ and a $25\,\si{\nano\second}$ gate time. The pulse parameters can be found in the caption of Fig.\ \ref{fig:app-single-gate}.

\section{Q-Q coupling}\label{app:two-gate}

Two coupled DSFQs are shown in Fig. \ref{fig:twoqubitgate}(a). The Lagrangian for the total circuit is
\begin{align}
    \mathcal{L} &= \frac{C}{2} \dot{\phi}_1^2 + \frac{C}{2} \dot{\phi}_2^2 + \frac{C}{2} \dot{\phi}_3^2 + \frac{C}{2} \dot{\phi}_4^2  \nonumber\\ 
    &+ \frac{C_{g}}{2} \qty(\dot{\phi}_1 - \dot{\phi}_3)^2 \nonumber\\
    &+ E_J \cos \phi_1 + E_J \cos \phi_2 + \alpha_1 E_J\cos\qty(\phi_1 - \phi_2 + \phiexta) \nonumber\\
    &+ E_J \cos \phi_3 + E_J \cos \phi_4 + \alpha_2 E_J \cos\qty(\phi_3 - \phi_4 + \phiextb).
\end{align}
By performing a Legendre transformation, we arrive at the result
\begin{align}
    H_{1(2)} &= 4E_C \qty(\frac{C + C_g}{C + 2C_g}) n_{1(3)}^2 + 4E_C n_{2(4)}^2 \nonumber \\ 
    &-E_J \cos \phi_{1(3)} + E_J \cos \phi_{2(4)} \nonumber\\ 
    &+ \alpha_{1(2)} E_J\cos\qty(\phi_{1(3)} - \phi_{2(4)} + \phi_{\mathrm{ext}_{1(2)}}),\nonumber\\
    H_{\mathrm{Q-Q}} &= 4E_C \qty(\frac{C_g}{C + C_g}) n_1n_3, \label{eq:H_two_qu}
\end{align}
where 1(2) refers to qubit 1(2) with charge and phase operators $n_{1(3)}, n_{2(4)}, \phi_{1(3)}, \phi_{2(4)}$.
The full Hamiltonian is a sum the two qubit Hamiltonians and the interaction term, $H=H_1 + H_2 + H_{\mathrm{Q-Q}}$. The qubit Hamiltonians have been renormalized due to the coupling capacitance between the two circuits. In Fig.\ \ref{fig:appendix_coupling}, we show the $\sigma_z^{(1)}\sigma
_z^{(2)}$ coupling due to the capacitive coupling defined by $\zeta_{ZZ} = \omega_{00} - \omega_{01} - \omega_{10} + \omega_{11}$. In Fig.\ \ref{fig:appendix_coupling}, it is apparent that the $\sigma_z^{(1)}\sigma
_z^{(2)}$ coupling is suppressed unless both barriers are lowered. Thus, single qubit gates where only one barrier is lowered do not give rise to unwanted $\sigma_z^{(1)}\sigma
_z^{(2)}$ interactions. \rettelse{However, we are limited to only half-grid single qubit gates if we neglect the next nearest neighbor stray capacitances.} \rettelse{As a final remark, we would like to point to the half-circular suppression of $\zeta_\mathrm{<<}$ in Fig.\ \ref{fig:appendix_coupling}. This interesting feature appears when the sign of the 
 $\sigma_z^{(1)}\sigma
_z^{(2)}$ interaction changes. In colloquial terms, the $\sigma_z^{(1)}\sigma
_z^{(2)}$ interaction is exactly cancelled when the ``push'' or ``pull'' on the $\ket{11}$ state from states below it is exactly compensated for by the push/pull from states above it.}
\bibliography{SCQubit}

\end{document}